\documentclass[aps,twocolumn,showpacs,amsmath,amssymb,floatfix,superscriptaddress,citeautoscript]{revtex4-1}
\usepackage{graphicx}
\usepackage{dcolumn}
\usepackage{bm}
\usepackage{amsmath}
\usepackage{amssymb,bbold}
\usepackage{verbatim}
\usepackage[normal]{subfigure}
\usepackage[bookmarks=false,linkcolor=blue,urlcolor=blue,colorlinks,citecolor=blue]{hyperref}

\begin{document}

\title{Topological superconductivity in full shell proximitized nanowires}

\author{Roman M. Lutchyn} \affiliation{Microsoft Quantum, Microsoft Station Q,
University of California, Santa Barbara, California 93106-6105 USA}

\author{Georg W. Winkler} \affiliation{Microsoft Quantum, Microsoft Station Q,
University of California, Santa Barbara, California 93106-6105 USA}

\author{Bernard van Heck} \affiliation{Microsoft Quantum, Microsoft Station Q,
University of California, Santa Barbara, California 93106-6105 USA}

\author{Torsten Karzig} \affiliation{Microsoft Quantum, Microsoft Station Q,
University of California, Santa Barbara, California 93106-6105 USA}

\author{Karsten Flensberg} \affiliation{Center for Quantum Devices, Niels Bohr Institute, University of Copenhagen, DK-2100 Copenhagen, Denmark}

\author{Leonid I. Glazman} \affiliation{Departments of Physics and Applied Physics, Yale University, New Haven, CT 06520, USA}

\author{Chetan Nayak} \affiliation{Microsoft Quantum, Microsoft Station Q,
University of California, Santa Barbara, California 93106-6105 USA}

\date{\today}

\begin{abstract}
We consider a new model system supporting Majorana zero modes based
on semiconductor nanowires with a full superconducting shell.  We demonstrate that, in the presence of spin-orbit coupling in the semiconductor induced by a radial electric field, the winding of the superconducting order parameter leads to a topological phase supporting Majorana zero modes. The topological phase persists over a large range of chemical potentials and can be induced by a predictable and weak magnetic field piercing the cylinder. The system can be readily realized in semiconductor nanowires covered by a full superconducting shell \cite{sole2018}, opening a pathway for realizing topological quantum computing proposals.
\end{abstract}

\maketitle

Majorana zero modes (MZMs) hold the promise to revolutionize quantum computation through topological quantum information processing \cite{Brouwer_Science, Nayak08, DasSarma2015}.
In the last decade, research in MZMs showed astonishing progress \cite{Mourik12,Deng12,Churchill13,Das12,Finck12,Nadj-Perge14,Ruby15,Albrecht16,Deng2016, Nichele2017,Gul18,Zhang2017}, fueled by proposals of simple and experimentally viable systems \cite{Fu08, Sau2010, Lutchyn10, Oreg10}.
In particular, existing routes towards realizing MZMs in semiconducting nanowires~\cite{Lutchyn10,Oreg10} rely on rather basic ingredients: spin-orbit coupling, a Zeeman field, and induced superconductivity.
Nevertheless, the required coexistence of large ($\sim$1T) magnetic
fields with superconductivity, as well as the need for careful control
of the chemical potential in the semiconductor, pose important
challenges towards a consistent realization of MZMs in nanofabricated
devices, requiring ongoing experimental improvements~\cite{Lutchyn17}.

\begin{figure}[htp]
  \begin{center}
    \includegraphics[width=\columnwidth]{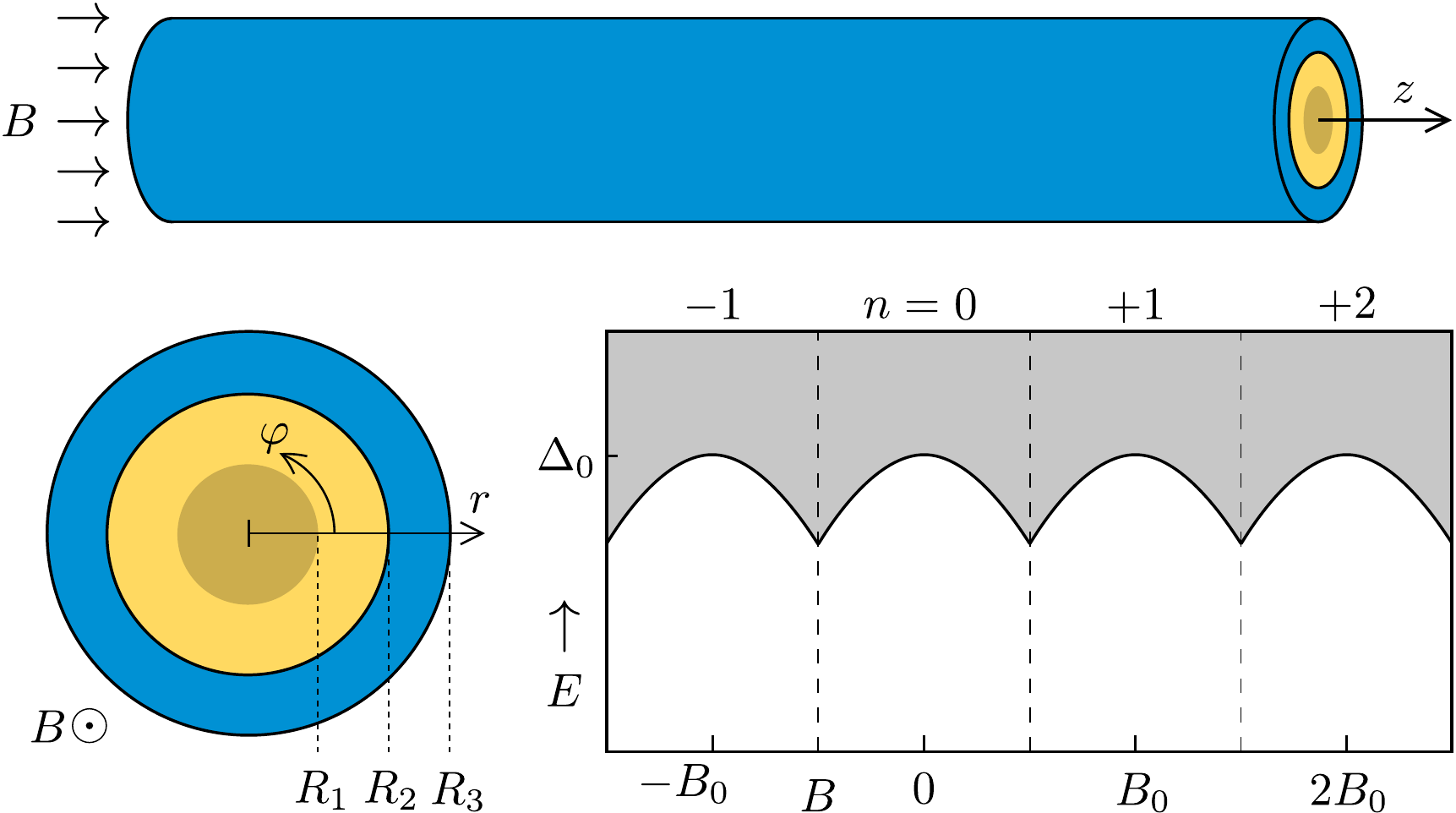}
    \caption{\emph{Top:} Illustration of a semiconducting nanowire (yellow) with a full superconducting shell (blue), subject to a weak axial magnetic field $B$. \emph{Bottom left}: Detail of the cross-section. The shaded yellow region with $r<R_1$ indicates the possible presence of an insulating core in the semiconductor. \emph{Bottom right}: Sketch of the energy gap in the superconducting shell as a function of the magnetic field, exhibiting characteristic Little-Parks lobes. Different lobes correspond to different winding numbers $n$ of the superconducting order parameter around the wire. The period is $B_0\approx 4\Phi_0/\pi(R_2+R_3)^2$ with $\Phi_0=h/2e$ the superconducting flux quantum.
      \label{fig:layout}}
  \end{center}
\end{figure}

In this Article, we show that a superconducting cylinder filled with a semiconducting core is an ideal alternative candidate for creating MZMs.
While being of similar simplicity and practical feasibility
\cite{Krogstrup15} as the original nanowire proposals~\cite{Lutchyn10,Oreg10}, full-shell
nanowires provide key advantages. 
First and foremost, the topological transition in a full-shell wire is driven by the field-induced winding of the superconducting order parameter, rather than by the Zeeman effect, and so the required magnetic fields can be very low ($\sim 0.1$T).
Therefore, the present proposal is compatible with conventional superconducting electronics and removes the need for a large $g$-factor semiconductor, potentially expanding the landscape of candidate materials.
Moreover, the full shell naturally protects the semiconductor from impurities and random surface doping, thus enabling a reproducible way of growing many wires with essentially identical electrostatic environments.
Although full-shell wires do not allow for direct  gating of the electron density in the semiconducting core, we demonstrate below that via a careful design of the wire properties, e.g. by choosing the right radius, it is possible to obtain wires that naturally harbor MZMs at a predictable magnetic field.

While it is known that well-chosen superconducting phase differences can be used to break time-reversal symmetry and localize MZMs in semiconductor heterostructures~\cite{Romito2012,vanHeck14,Kotetes2015,Hell2017,Pientka17,Stanescu18}, the corresponding realizations typically require careful tuning of the fluxes which would complicate a scalable approach with multiple MZMs \cite{Karzig16}.
Here, we show that the quantized superconducting winding number in a full-shell wire is a natural and more robust implementation of the wanted phase differences, leading to sizable regions of topological phase space.
Unlike previous works based on vortex lines in proximitized topological insulators \cite{Hosur11,Chiu11,Cook11}, our proposal does not rely on bulk topological properties of the host materials, but can be realized with conventional semiconductors.
We first demonstrate our ideas in a simple model of a hollow semiconducting core, where an analytic mapping to the standard model of a topological superconductor~\cite{Lutchyn10,Oreg10} is possible. 
We complement these results with numerical and analytical studies of the topological phase in the opposite regime where the electron density is spread out over the entire semiconducting core.

{\it Theoretical model.}---We consider a nanowire consisting of a
semiconducting core and a full superconducting shell, see
Fig.~\ref{fig:layout}.  We assume that the semiconductor (e.g., InAs)
has a large Rashba spin-orbit coupling due to an intrinsic electric
field pointing in the radial direction at the semiconductor-superconductor interface.  The system is subject to a magnetic
field along the direction of the nanowire $\hat{z}$, i.e.
$\vec{B}=B \hat{z}$.  Using cylindrical coordinates and the symmetric
gauge for the electromagnetic vector potential,
$\vec{A}=\frac{1}{2} (\vec{B} \times \vec{r})$, the effective
Hamiltonian for the semiconducting core can be written as (henceforth
$\hbar=1$)
\begin{align}
H_0=\frac{(\vec{p}+e A_{\varphi}\hat{\varphi})^2}{2m^*}-\mu+\alpha\, \hat{r}\cdot \left[\vec{\sigma} \times  (\vec{p}+e A_{\varphi}\hat{\varphi})\right].
\end{align}
Here $\vec{p}$ is the electron momentum operator, $e>0$ the electric charge,
$m^*$ the effective mass, $\mu$ is the
chemical potential, $\alpha$ the strength of the Rashba spin-orbit
coupling, and finally $\sigma_i$ are spin-$\frac{1}{2}$ Pauli
matrices. $\hat{r}$, $\hat{\varphi}$ and $\hat{z}$ are the cylindrical unit vectors. For ease of presentation, we consider $r$-independent $\mu$ and $\alpha$ in our effective model, which may be viewed as averaged versions of the corresponding $r$-dependent quantities.  The vector potential $A_{\varphi}= \Phi(r)/2\pi r$, where $\Phi(r)=\pi B r^2$ is the flux threading the cross-section at radius $r$.
For simplicity, we neglect the Zeeman term due to the
small magnetic fields required in these devices.

The shell (e.g., made out of Al) induces superconducting correlations in the nanowire due to Andreev processes at the semiconductor-superconductor interface.
If the coupling to the superconductor is weak, the induced pairing in the nanowire can be expressed as a local potential $\Delta(\vec{r})$ (see Appendix~\ref{app:effective}).
In the Nambu basis $\Psi=(\psi_{\uparrow}, \psi_{\downarrow},\psi^\dag_{\downarrow}, -
\psi^\dag_{\uparrow})$, the Bogoliubov-de-Gennes (BdG) Hamiltonian for
the proximitized nanowire is then given by
\begin{align}\label{eq:HBdG}
H_{\rm BdG}=
\left[\begin{array}{cc}
H_0(\vec{A}) & \Delta(\vec{r}) \\
\Delta^*(\vec{r}) & -\sigma_y H_0(-\vec{A})^* \sigma_y
\end{array}\right].
\end{align}
We assume that the thickness of the SC shell is smaller than London penetration depth: $R_3-R_2\ll\lambda_L$.
Therefore, the magnetic flux threading the SC shell is not quantized.
However, the magnetic field induces a winding of the superconducting phase, i.e.~the order parameter $\Delta(\vec{r})=\Delta(r) e^{-i n\varphi}$ with $\varphi$ the angular
coordinate and $n\,\in\,\mathbb{Z}$ the winding number.
In practice, the winding number $n$ adjusts itself to the value of the external magnetic field so that the free energy of the superconducting shell is minimized.
This is the familiar Little-Parks effect \cite{little-parks}: the changes in winding number lead to periodic lobes in the energy spectrum of the superconducting shell, see Fig.~\ref{fig:layout} and  Fig.~S1.

We notice the following rotational symmetry of the BdG Hamiltonian: $[J_z,H_{\rm BdG}]=0$ with $J_z=-i \partial_{\varphi}+\frac{1}{2}\sigma_z+\frac{1}{2}n \tau_z$,
where we introduced $\tau_i$ matrices acting in Nambu space.
Eigenstates of $H_\textrm{BdG}$ can thus be labeled by a conserved quantum number $m_J$:
$  \Psi_{m_J}(r,\varphi, z)\propto e^{i \left(m_J-\frac{1}{2}\sigma_z-\frac{1}{2}n \tau_z\right)\varphi} \Psi_{m_J}(r, z)$.
The wave function has to be single-valued, which imposes the following
constraint on $m_J$:
\begin{align}
m_J \in \begin{cases}
\mathbb{Z} &  n\;{\rm odd}\,,\\
\mathbb{Z}+\frac{1}{2} & n\;{\rm even}\,.
\end{cases}
\end{align}
We remove the angular dependence of $H_{\rm BdG}$ via a unitary
transformation $U=\exp\left[ -i \left(m_J-\frac{1}{2}\sigma_z-\frac{1}{2}n \tau_z\right)\varphi \right]$, namely $\tilde{H}_{\rm BdG}=UH_{\rm BdG}U^\dag$ where
\begin{align}\label{eq:HBdGr}
\tilde{H}_{\rm BdG}&=\left(\frac{p_{z}^2}{2m^*}+\frac{p_r^2}{2m^*}-\mu\right)\tau_z \\\nonumber
&+\frac{1}{2m^*r^2}\left(m_J-\frac{1}{2}\sigma_z-\frac{1}{2}n\tau_z + eA_\varphi r\tau_z\right)^2\tau_z \\ \nonumber
&- \frac{\alpha}{r}\sigma_z \tau_z\left(m_J-\frac{1}{2}\sigma_z-\frac{1}{2}n \tau_z + eA_\varphi r\tau_z\right)\\
& +\alpha p_z \sigma_y \tau_z+\Delta(r)\tau_x.\nonumber
\end{align}
Here $p_r^2=-\frac{1}{r}\frac{\partial}{\partial r}r\frac{\partial}{\partial r}$ and $p_z=-i \frac{\partial}{\partial z}$. Note that naively one would expect spin-orbit coupling to average
out. However, the non-trivial structure of $m_J$-eigenvectors yields
finite matrix elements proportional to the Rashba spin-orbit coupling. 
We will now show that the above BdG Hamiltonian supports
topological superconductivity and MZMs.

\begin{figure}[htp]
  \begin{center}
    \includegraphics[width=\columnwidth]{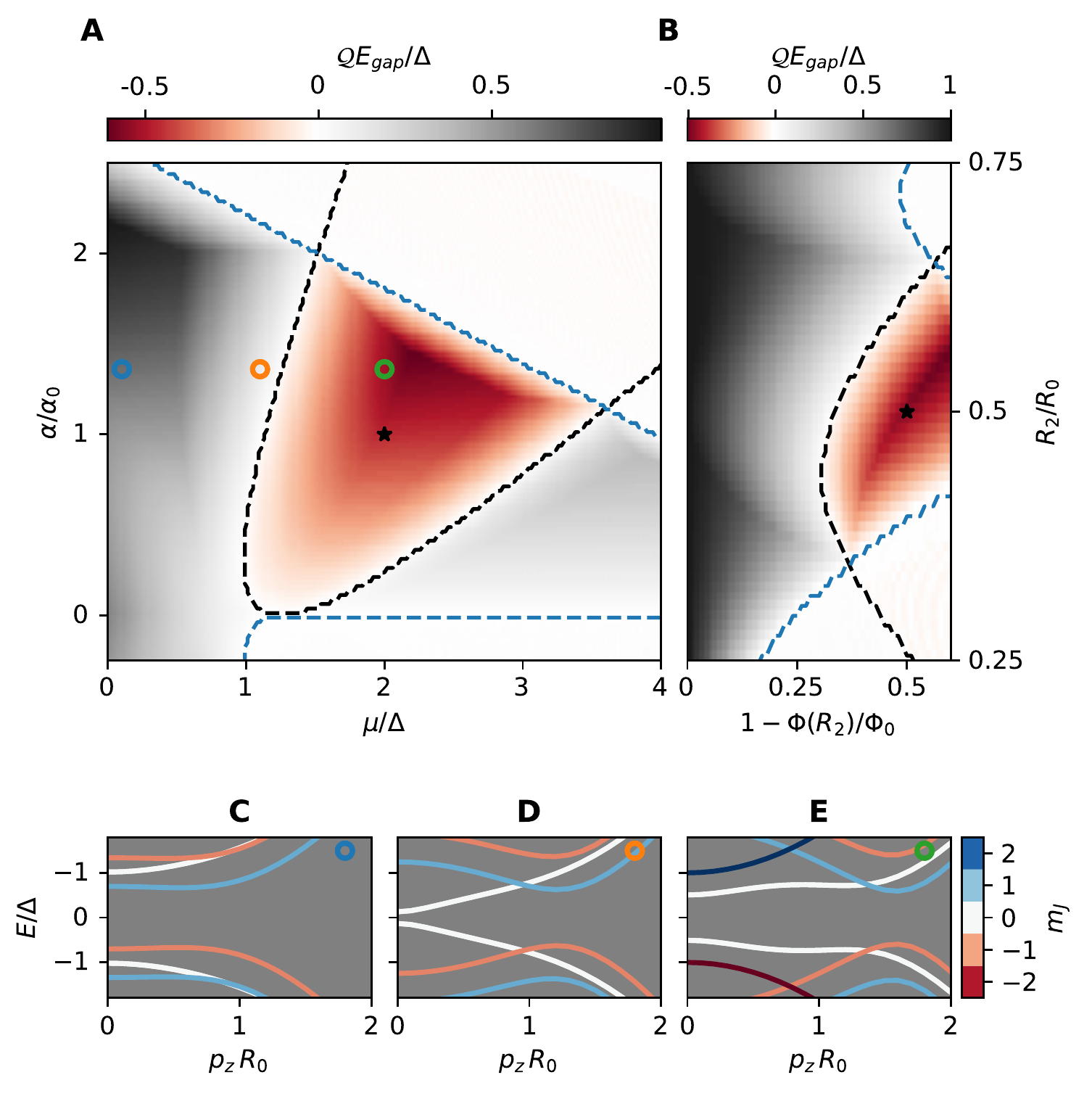}
\caption{Topological phase diagram of the hollow-cylinder model. \textbf{A}, Bulk energy gap $E_\textrm{gap}$ as a function of chemical potential and spin-orbit coupling. The energy gap is multiplied by the topological index $\mathcal{Q}=\pm 1$, so that red regions correspond to the gapped topological phase. The black dashed line denotes the boundary of the topological phase in the $m_J=0$ sector, according to Eq.~\eqref{eq:topological_condition}, while the blue dashed lines denote the boundaries at which higher $m_J$ sectors become gapless, see Appendix~\ref{app:highermJ}. Here $\Phi(R_2)/\Phi_0=\tfrac{1}{2}$, $R/R_0=\tfrac{1}{2}$, as indicated by a black marker in \textbf{B}. We define $\alpha_0=\sqrt{\Delta/2m}$ and $R_0=1/\sqrt{2m\Delta}$. For reference, using realistic parameters $m^*=0.026\;m_\textrm{e}$ and $\Delta=0.2$~meV, one obtains $\alpha_0\approx 17$~meV$\cdot$nm and $R_0\approx 85$~nm. \textbf{B}, Bulk energy gap at fixed $\mu/\Delta=2$ and $\alpha/\alpha_0=1$, as indicated by a black marker in \textbf{A}, as a function of flux and $R$. \textbf{C}-\textbf{E} Band-structures at the points indicated in \textbf{A}, illustrating the closing and re-opening of the bulk gap in the $m_J=0$ sector.
  \label{fig:thin_shell}}
\end{center}
\end{figure}

{\it Hollow cylinder model.}---We now focus on the limit in which the
semiconductor forms a thin-wall hollow cylinder (i.e., $R_1 \approx R_2$ in
Fig.~\ref{fig:layout}).  This approximation is motivated by the fact
that there is an accumulation layer in certain
semiconductor-superconductor heterostructures such as InAs/Al, so that
the electron density is concentrated within the screening length (typically $\sim 20-30$~nm) from the interface~\cite{Antipov2018, Mikkelsen2018}. In this case, one can consider only the lowest-energy radial mode in Eq.~\eqref{eq:HBdGr}. This allows for an analytical solution of the model. The effective Hamiltonian for the hollow-cylinder model reads
\begin{align}\label{eq:H_mj}\nonumber
\tilde{H}_{m_J}&=
\left[\frac{p_{z}^2}{2m^*}-\mu_{m_J}\right]\!\tau_z+V_{Z}\sigma_z\\
&+A_{m_J} + C_{m_J}\sigma_z\tau_z+\alpha p_z \sigma_y \tau_z+\Delta\tau_x.
\end{align}
Here, $\Delta\equiv\Delta(R_2)$ and the parameters $\mu_{m_J}$ and $V_{Z}$ correspond to the effective chemical potential and Zeeman energy. $A_{m_J}$ and $C_{m_J}$ represent the coupling of the generalized angular momentum $J_z$ with magnetic field and electron spin, respectively. They are defined as  
\begin{align}\label{eq:effparam}
\mu_{m_J}&=\mu-\frac{1}{8m^*R_2^2}\left(4m_J^2+1+\phi^2\right)-\frac{\alpha}{2R_2}\,,\\
V_{Z}&=\phi\,\left(\frac{1}{4m^*R_2^2}+\frac{\alpha}{2R_2}\right)\,,\\
A_{m_J}&=-\frac{\phi m_J}{2m^* R_2^2}\,,\\
C_{m_J}&=-m_J\left(\frac{1}{2m^*R_2^2}+\frac{\alpha}{R_2}\right)\,,
\end{align}
with $\phi=n-\Phi(R_2)/\Phi_0$.
Particle-hole symmetry relates states with opposite energy and angular quantum number $m_J$, i.e., $\mathcal{P}\Psi_{E, m_J}=\Psi_{-E, -m_J}$ with $\mathcal{P}=\tau_y \sigma_y \mathcal{K}$, where $\mathcal{K}$ represents complex conjugation.
Thus, the $m_J=0$ sector, which is allowed when the winding number $n$ is {\it odd}, is special as it allows non-degenerate Majorana solutions at zero energy.

Let us consider the $m_J=0$ sector and $n=1$.
In this case, $A_{0}=0$ and $C_{0}=0$, and one can map Eq.~\eqref{eq:H_mj} to the Majorana nanowire model of Refs.~\cite{Lutchyn10, Oreg10}.
Notice that the effective Zeeman term has an orbital origin here and is present even when the g-factor in the semiconductor is zero. Both $\mu_0$ and $V_Z$ can be tuned by the magnetic flux $\Phi(R_2)$, which may induce a topological phase transition.
When the core is penetrated by one flux quantum (i.e., $\Phi(R_2)=\Phi_0$), then $V_Z=0$. This regime corresponds to the trivial (s-wave) superconducting phase.
However, a small deviation of the magnetic field can drive the system into the topological superconducting phase~\footnote{Note that magnetic flux piercing the finite-thickness superconducting shell can be significantly different from that penetrating the core.}.
Indeed, the Zeeman and spin-orbit terms in Eq.~\eqref{eq:H_mj} do not commute
and thus $V_Z$ opens a gap in the spectrum at $p_z=0$.  At the topological quantum phase transition between the two phases, the gap in the $m_J=0$ sector,
\begin{align}\label{eq:topological_condition}
E_\text{gap}^{(0)}=\left||V_Z|-\sqrt{\mu_{0}^2+\Delta^2}\right|\,,
\end{align}
closes. The resulting phase diagram is shown in Fig.~\ref{fig:thin_shell},
where the gap closing at the topological transition is indicated by black dashed lines. Close to the transition, the quasiparticle spectrum in the $m_J=0$ sector is given by 
\begin{align}\label{eq:gap}
\!E(p_z)=\sqrt{\Big(E_\text{gap}^{(0)}\Big)^2+(vp_z)^2}. 
\end{align}
with $v=\alpha \Delta/\sqrt{\Delta^2+\mu_0^2}$ and corresponding coherence length $\xi\sim v/E_\text{gap}^{(0)}$.    

We now consider the effect of $m_J\neq 0$ sectors. In general, the topological phase diagram can be obtained by calculating the topological index ${\cal Q}$~\cite{Kitaev01},
\begin{align}
{\cal Q}=\mbox{\rm sign}\prod_{m_J \in Z}\left[\Delta^2+(C_{mj}-\mu_{mj})^2-(A_{mj}+V_Z)^2\right], \label{indexQ}
\end{align}
where ${\cal Q}=1$ and ${\cal Q}=-1$ correspond to trivial and
topological phases, respectively. A well-defined topological phase requires the quasiparticle bulk gap to be finite for all values of $m_J$. Due to the angular symmetry of Eq.~\eqref{eq:H_mj}, different $m_J$ sectors do not mix and, as a result, the condition for a finite gap in the $m_J\neq 0$ sectors is $ \Delta^2+ (C_{mj}-\mu_{mj})^2 > (A_{mj}+V_Z)^2$, see Appendix~\ref{app:highermJ}. Thus, the topological phase supporting MZMs appears due to the change of ${\cal Q}$ in the $m_J=0$ sector. In Fig.~\ref{fig:thin_shell} we show the topological phase diagram and energy gap of the hollow cylinder model determined by taking into account all $m_J$ sectors. The above-mentioned conditions restrict the extent of the gapped topological phase to small chemical potentials and spin-orbit couplings. Nevertheless, Fig.~\ref{fig:thin_shell} demonstrates that in the hollow-cylinder model such phase exists over an extended range in all the model parameters, with optimal quasiparticle gaps comparable in magnitude to $\Delta$ and a corresponding coherence length $\xi \sim 100$ nm.

\begin{figure}
  \begin{center}
    \includegraphics[width=\columnwidth]{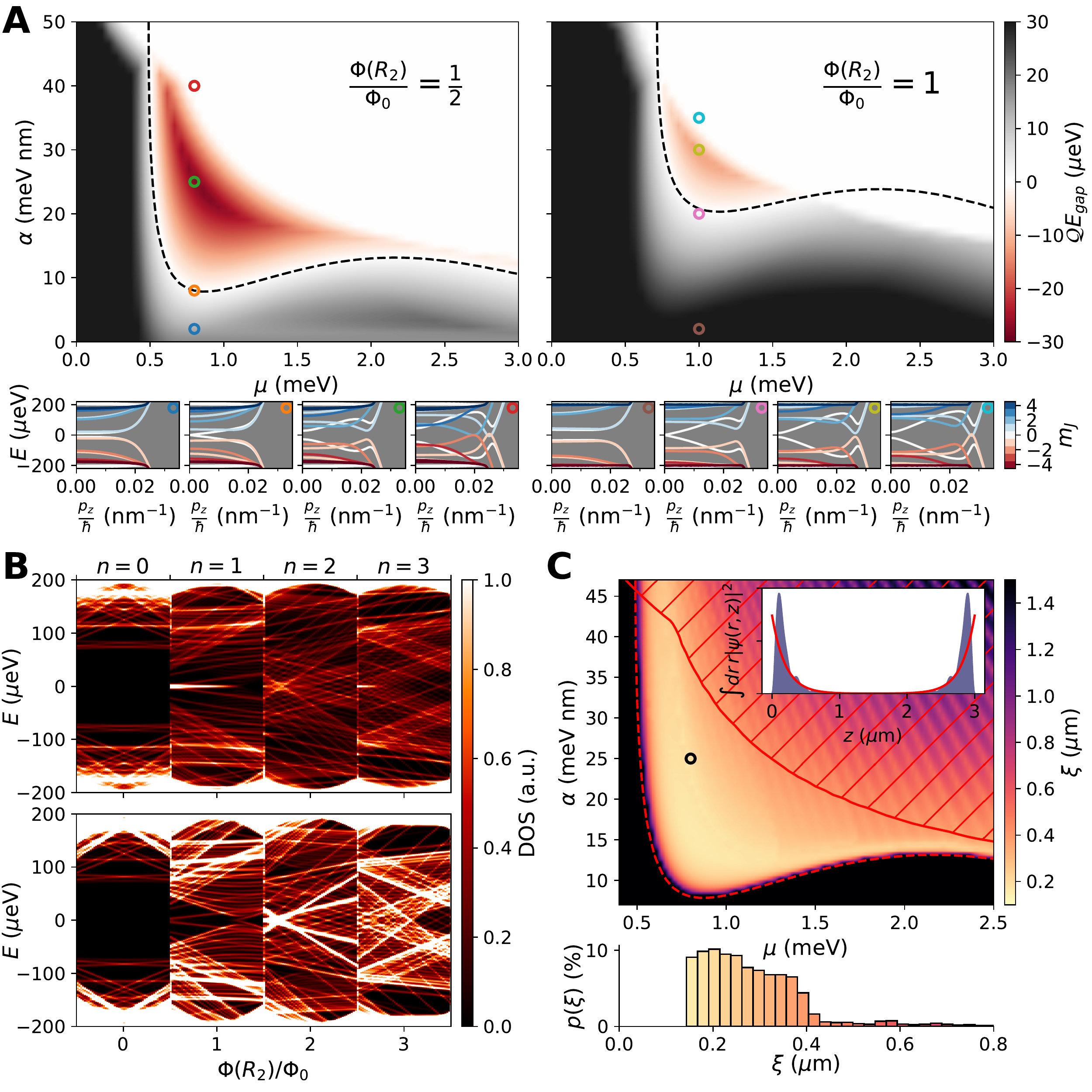}
\caption{{\bf A}, Topological phase diagram for the full-cylinder model with
  $R_2=100$\,nm, $m^*=0.026\,m_e$, 
  $\Delta=0.2$\,meV, and $\Phi(R_2)=\Phi_0/2$ (left panel) and $\Phi(R_2)=\Phi_0$ (right panel). The bulk energy gap
  $E_\textrm{gap}$, multiplied by the topological index $\mathcal{Q}$,
  is plotted as a function of $\mu$ and $\alpha$.
  The dashed line denotes the boundary of the topological phase 
  obtained by finding the zero energy crossing at $p_z=0$ in the
  $m_J=0$ sector. Below we plot band structures at the points indicated
  in the phase diagrams above. The color of the bands indicate the $m_J$ sector. 
  {\bf B}, DOS at the end (top panel) and in the middle (bottom panel) of a finite wire of 3\,$\mu$m length as a function of flux for $\mu=1.1$\,meV, 
  $\alpha=30$\,meV\,nm and the same parameters as in {\bf A}.
  {\bf C}, Majorana coherence length for the full-cylinder model with
  the same parameters as in {\bf A} using $\Phi(R_2)=\Phi_0/2$.  The coherence length is
  obtained by fitting the exponential decay of the Majorana
  wavefunction in a wire of 3\,$\mu$m length. An example fit is presented in the inset showing the Majorana wavefunction integrated over radius for $\mu=0.8$\,meV and
  $\alpha=25$\,meV\,nm as indicated by the black circle.
  The dashed line denotes the boundary of the topological phase in
  the $m_J=0$ sector and the shaded region is gapless due to higher $m_J$
  sectors.  The histogram below shows the probability of finding a coherence length in the interval given by the width of the bin 
  inside of the gapped topological phase.
  \label{fig:full_core}}
\end{center}
\end{figure}

\paragraph*{Full-cylinder model.} 
Now we consider the case in which the
electron density is uniform in the semiconducting core (i.e., $R_1=0$
in Fig.~\ref{fig:layout}).
We solve for the radial modes in the core numerically for the different $m_J$ quantum
numbers~\cite{Winkler17}.
The superconductor is treated effectively as a boundary condition at $r=R_2$, neglecting the effect of the magnetic field penetrating the shell, so that $\Phi(R_3)\approx\Phi(R_2)$.
This treatment of the proximity effect is justified for a thin superconducting shell in the dirty limit~\cite{Larkin1965} (see Appendix~\ref{app:shell},~\ref{app:effective} and~\ref{app:clean_numerics_details} for technical details).

In Fig.~\ref{fig:full_core}~{\bf A} we show the topological phase diagram for the full-cylinder model with parameters appropriate for InAs-Al hybrid semi-superconductor nanowires.
As in the hollow-cylinder model, one finds a stable topological phase which extends over a reasonably large range of the chemical potential and has maximum topological gap of order $30$\,$\mu$eV.
A large part of the topological phase is gapless due to $m_J \ne 0$ states as in the previous case.
Due to the large extent of the radial wavefunction into the semiconducting core the topological gap is smaller in the full-cylinder model. Also, the cancellation of the superconducting winding by the orbital effect is not exact in this case, so that a topological phase also appears at $\Phi(R_2)=\Phi_0$ for appropriate parameters, see also the analytical solution in the Appendix~\ref{app:full_cylinder_analytics}. In Fig.~\ref{fig:full_core}~{\bf A} we show the momentum
dispersion of different $m_J$ sectors, illustrating the topological
transition in and out of the gapped topological phase.  The bands
forming in the core of the wire have a distinctly flat dispersion as can be seen in Fig.~\ref{fig:full_core}~{\bf A},
which is reminiscent of Caroli-de Gennes-Matricon vortex states~\cite{caroli1964bound}.

The evolution of the local density of states (DOS) at the end of a finite wire as a
function of magnetic field is shown in Fig.~\ref{fig:full_core}~{\bf B}. As the flux $\Phi(R_2)$ is increased, the
winding number $n$ changes by one at every half-integer multiple of
$\Phi_0$.  The change in winding number causes a discontinuous jump in
the density of states.
At energies close to the pairing gap, the DOS reproduces the periodic Little-Parks lobes already sketched in Fig.~\ref{fig:layout}.
However, at lower energies, the DOS reveals the sub-gap spectrum in the semiconducting core.
A peak in the DOS is visible at zero energy in the $n=1$
and $n=3$ lobe, but not in the $n=0$ and $n=2$ lobes, in agreement
with the fact that isolated MZMs should only appear for odd
values of $n$. Within odd lobes one can see the characteristic asymmetry of the subgap spectra with respect to the center of the lobe which stems from the difference in magnetic flux penetrating the core of the wire and the superconducting shell. The disappearance of the Majorana zero-energy states within odd lobes occurs because both $\mu_0$ and $V_Z$ depend on the magnetic field, see Eqs.~\eqref{eq:effparam} which, at some point, leads to a topological phase transition. The bulk gap closing  at around $\Phi(R_2)/\Phi_0\approx 1.25$ is clearly visible when plotting the DOS in the middle of the wire as shown in the lower panel of Fig.~\ref{fig:full_core}~{\bf B}.

In Fig.~\ref{fig:full_core}~{\bf C}, we investigate Majorana hybridization due to a finite nanowire length ($L = 3\mu$m) and extract the coherence length by fitting the Majorana wave function, see the inset. 
Despite the relatively small topological gaps of Fig.~\ref{fig:full_core}~{\bf A}, we find quite short Majorana coherence lengths with the minimum being of the order of 160\,nm due to the small group velocity of the bulk states of Fig.~\ref{fig:full_core}~{\bf A}. 

\begin{figure}
  \begin{center}
    \includegraphics[width=\columnwidth]{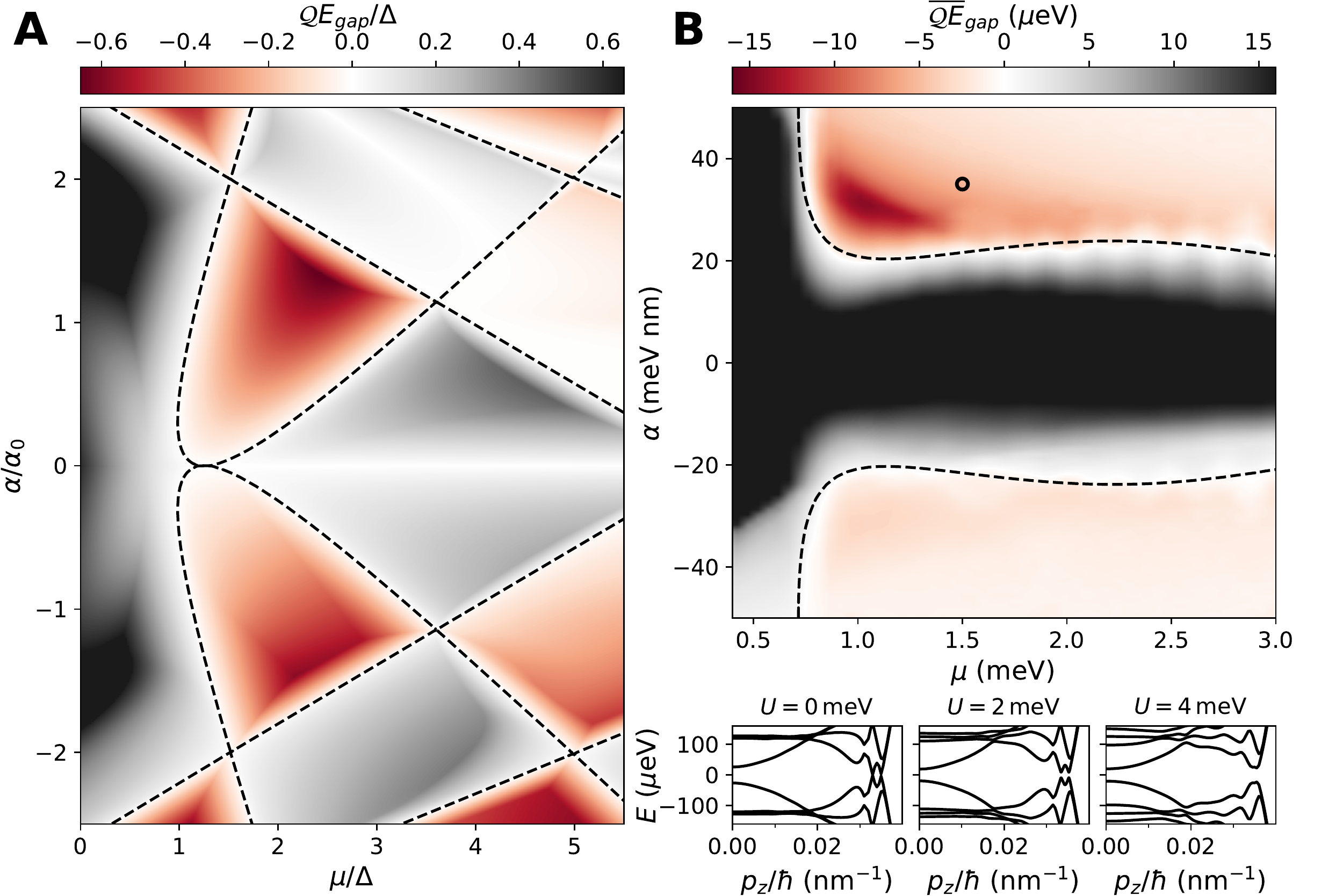}
\caption{
{\bf A}, Topological phase diagram for the hollow cylinder model with broken angular symmetry. The symmetry is broken by introducing an anisotropic Rashba spin-orbit coupling (see also Appendix~\ref{app:highermJ}) of which the $z$-component $\alpha_2$ is dependent on $\varphi$: $\alpha_2 = \alpha \left(1 + q \cos(2\varphi)\right)$.
We show results for $q=1$, $\Phi(R_2)/\Phi_0=\tfrac{1}{2}$ and $R/R_0=\frac{1}{2}$.
{\bf B}, Topological phase diagram of the disordered full cylinder model
  with the same parameters as in
  Fig.~\ref{fig:full_core} using  $\Phi(R_2)=\Phi_0$. To allow for cylindrical
  symmetry breaking perturbations the
  Hamiltonian~\eqref{eq:HBdG} is discretized on a square
  lattice with $a=10$\,nm in the two
  dimensional cross section. The disorder potential
  $\delta U$ on each lattice site in the superconductor is uniformly distributed $\delta U
  \in [-U, U]$ with $U=2$\,meV. We consider rotational symmetry breaking disorder that is translationally invariant in the $z$-direction. The gap times the topological index
  is averaged over 12 realizations, see Appendix~\ref{app:disorder_numerics_details}. The dashed lines indicate the phase boundaries without
  disorder. 
   Below we show the band structures at $\alpha=35$\,eV\,nm and $\mu=1.5$\,meV for
  increasing disorder strength from left to right $U=\{0, 2,
  4\}$\,meV. A single disorder configuration is shown.
  \label{fig:disorder}} 
\end{center}
\end{figure}

\paragraph*{Effect of angular-symmetry-breaking perturbations.} 

We now investigate perturbations breaking the angular symmetry of the Hamiltonian~\eqref{eq:HBdGr}. Such perturbations (e.g., shape deformations or disorder in the superconducting shell) are ubiquitous in realistic devices and would couple different $m_J$ eigenstates. This may actually have beneficial consequences for the stability of the topological phase. 
Indeed, the perturbations that couple different $m_J$ sectors may open a gap in regions of large $\mu$ and $\alpha$, see
Fig.~\ref{fig:full_core}~{\bf A} which, otherwise, are gapless due to the closing of the gap at finite momentum. It is enlightening to consider angular-symmetry-breaking perturbations that are momentum-dependent and, in particular, such that they vanish at $p_z=0$ but open the gap at finite $p_z$, see Fig.~\ref{fig:disorder}~\textbf{A}. One such example would be an angular symmetry breaking spin-orbit coupling resulting from the electric fields in a full-shell nanowire with non-cylindrical geometry. In this case, the phase diagram can be obtained analytically for a hollow cylinder model using Eq.\eqref{indexQ} since such perturbations do not effect the gap closing at $p_z=0$. One may notice that the topological phase space now significantly increases and, in particular, now extends to negative $\alpha$, see Fig.~\ref{fig:disorder}~\textbf{A}.   

We have also numerically studied the effect of disorder in the superconductor within the full cylinder model, see Fig.~\ref{fig:disorder}. Disorder in the superconductor breaks angular symmetry and leads to a phase diagram consistent with the discussion above. Indeed, the topological phase now also appears at large $\mu$ and $|\alpha|$. This suggests that the topological phase may exist over a large parameter regime in physical wires, consistent with flux-induced zero-bias peaks in experiments on full-shell nanowires~\cite{sole2018}.

\paragraph*{Conclusions and Outlook.} In this Article we investigated a novel physical system supporting MZMs based on semiconductor nanowires covered by a superconducting shell~\cite{sole2018}. Using a combination of analytical and numerical methods, we calculated the topological phase diagram and showed that the model supports robust topological superconductivity in a reasonably large parameter space. We characterized the topological phase by calculating quasiparticle gap and effective coherence length. The existence of a readily accessible robust topological phase in full-shell nanowires opens a pathway for realizing topological quantum computing proposals.     

\section*{Acknowledgments}  We thank Saulius Vaitiek\.{e}nas, Ming-Tang Deng, Peter Krogstrup, Charlie Marcus and Michael Freedman for stimulating discussions. This work was performed in part at Aspen Center for Physics, which is supported by National Science Foundation grant PHY-1607611.
\appendix

\section{Model for the disordered superconducting shell.}
\label{app:shell}

\begin{figure}
  \begin{center}
    \includegraphics[width=\columnwidth]{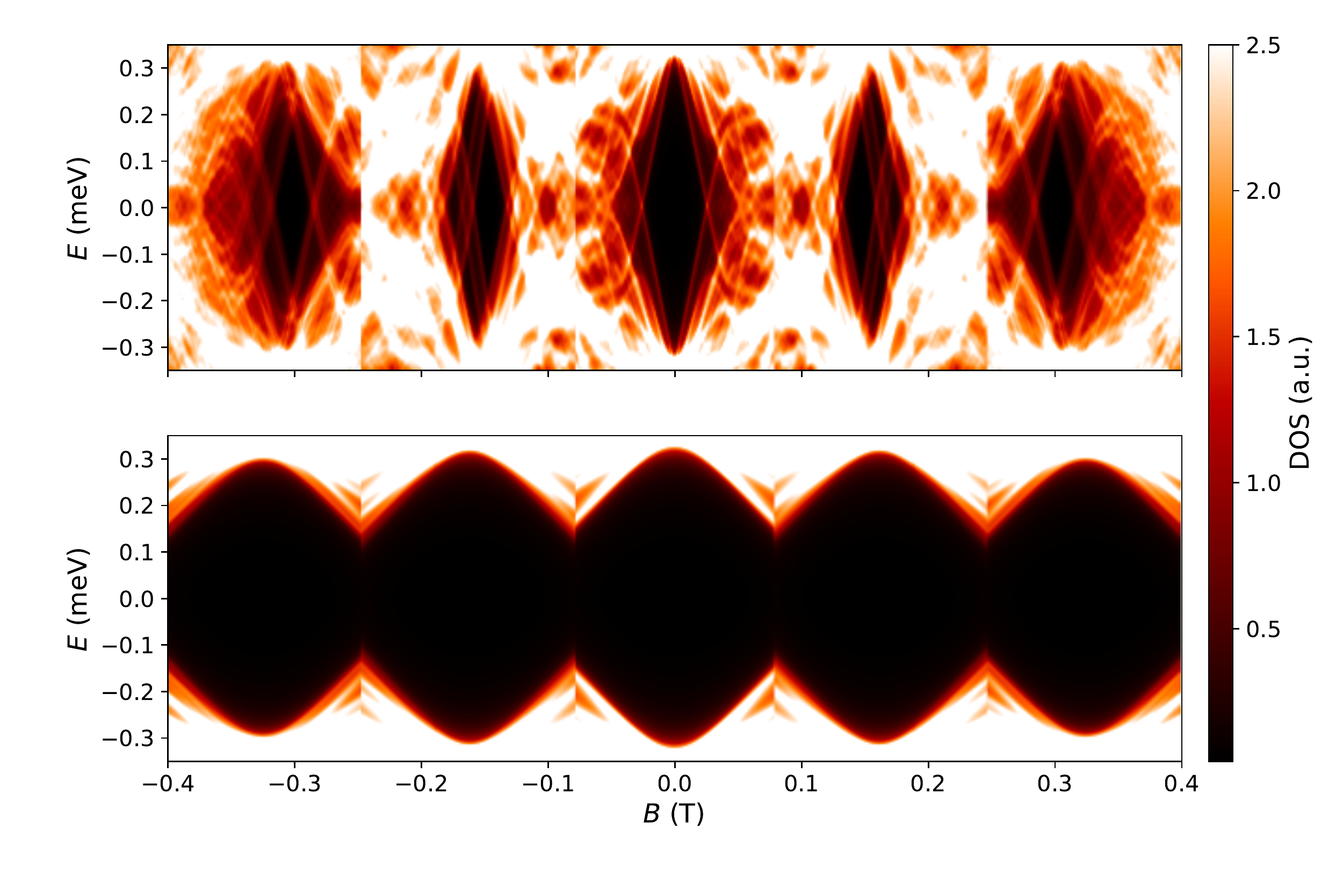}
\caption{We simulate a superconducting shell, without the semiconducting core, with $R_1 = R_2=60$\,nm and $R_3=70$\,nm. Realistic parameters corresponding to Al are used: $m^*=m_e$,  $\mu=10$\,eV and $\Delta_0=0.34\,$meV~\cite{Cochran1958}. The Hamiltonian Eq.~\eqref{eq:shamiltonian} is discretized on a square lattice with $a=0.1$\,nm using the Kwant package~\cite{kwant}. {\it Top:} We show the clean case, where the superconductivity is rapidly destroyed by the magnetic field when the radius $R_3$ is smaller than the coherence length $\xi_0$ in a clean superconductor. {\it Bottom:} We show the disordered case using the on-site disorder potential $\delta U$ which is randomly sampled from $\delta U
  \in [-U, U]$ with $U=2$\,eV. The disorder is applied in an outer layer of 5\,nm thickness, with the purpose of modelling an oxidized Al$_2$O$_3$ layer. The amplitude of Little-Parks oscillations is small because $R_3 > \sqrt{\xi_0 l}$ with $l$ being the mean-free path, see also Ref.~\cite{Schwiete2009}. \label{fig:realistic_superconducting_shell}}
\end{center}
\end{figure}

In this Section, we consider a disordered superconducting shell (e.g., Al shell) with inner and outer radii $R_2$ and $R_3$, respectively, see Fig.~1 of the main text.
We assume that the thickness of the shell  $d\equiv R_3-R_2 \ll \lambda_L$, with $\lambda_L$ being the London penetration length in the bulk superconductor.
In this case, the screening of the magnetic field by the superconductor is weak and can be neglected.
The effective Hamiltonian for the SC shell in cylindrical coordinates can be written as
\begin{align}
  \!\!H^{(s)}_{\rm BdG}\!&=\left[\frac{\hat{p}_{z}^2}{2m^*}\!+\!\frac{\hat{p}_{r}^2}{2m^*}\!+\!\frac{(\hat{p}_{\varphi}\!+\!eA_{\varphi}\tau_z)^2}{2m^*}\!-\!\mu^{(s)}+V_{\rm imp} \right] \tau_z\! \nonumber\\
                   &+\!\Delta_0\left[\cos(n \varphi)\tau_x\!+\!\sin(n \varphi)\tau_y\right]
                     \label{eq:shamiltonian}
\end{align}
Here, $\hat{p}_i$ are the electron momentum operators, $e>0$ the electric charge,
$m$ the electron mass in the SC, $A_{\varphi}=\frac{1}{2}Br$, $\mu^{(s)}$ is the
chemical potential in the SC, $\tau_i$ are Pauli matrices representing Nambu space, $\Delta_0$ is bulk SC gap at $B=0$, $n$ is the winding number for the SC phase, and $V_{\rm imp}$ represents short-range impurity scattering potential. It is enlightening to perform a gauge transformation which results in a real order parameter, i.e. $\Delta_0\left[\cos(n \varphi)\tau_x\!+\!\sin(n \varphi)\tau_y\right]\rightarrow \Delta_0 \tau_x$).
The gauge transformation introduces an effective vector potential, $A_{\varphi} \rightarrow \tilde{A}_{\varphi}$ with
\begin{equation}
\tilde{A}_{\varphi}=-\frac{1}{2er} (n-2eA_{\varphi}r)=-\frac{1}{2e r}\left[n-\frac{\Phi(r)}{\Phi_0}\right]
\end{equation}
where $\Phi(r)=\pi B r^2$ and $\Phi_0=h/2e$. It follows from this argument that the solution of Eq.~\eqref{eq:shamiltonian} should be periodic with $\Phi_0$, see Fig.~\ref{fig:realistic_superconducting_shell}.
Namely, the winding number adjusts itself to the value of the magnetic field so that the energy of the superconductor is minimized.
In particular, for each winding number $n$, the maxima of the quasiparticle gap occur at
\begin{equation}
B_n\approx 4n\frac{\Phi_0}{\pi (R_2+R_3)^2}\,.
\end{equation}
We neglect the Zeeman contribution since the typical magnetic fields of interest are smaller than $100$ mT for which the Zeeman splitting is negligible.

In order to understand the magnetic field dependence of the quasiparticle gap, one needs to calculate the Green's functions for the disordered SC shell as a function of $\tilde{A}_{\varphi}$.
The disordered superconductor is characterized by an elastic mean free path $l_e$ and a corresponding diffusive coherence length $\xi_d = \sqrt{l_e\,\xi_0} \gg\textbf{} l_e$, where $\xi_0=v_F/\Delta$ is the coherence length in the bulk, clean limit ($v_F$ is the Fermi velocity in the SC).
For simplicity, we assume henceforth that the thickness of the superconducting shell $d \gtrsim \xi_d$~\footnote{our results also apply to the case of $d \sim l_e$ and $\xi_d \gtrsim d$}, so that the properties of the system can be obtained by calculating the Green's function for the disordered bulk superconductor in magnetic field $B$ and $n=0$.
This problem was considered by Larkin~\cite{Larkin1965}, who showed that within the self-consistent Born approximation the normal and anomalous Matsubara Green's function are given by
\begin{align}
G^{(m_J)}(\omega_n, \varepsilon)= \frac{i\omega_n+\bar{G}+H m_J}{(\Delta+\bar{F})^2+\varepsilon^2-(i\omega_n+\bar{G}+H m_J)^2}\\
F^{(m_J)}(\omega_n,\varepsilon)=-\frac{\Delta+\bar{F}}{(\Delta+\bar{F})^2+\varepsilon^2-(i\omega_n+\bar{G}+H m_J)^2}
\end{align}
where $H=eB/4m$ and $m_J$ is the angular momentum eigenvalue and $\epsilon$ is the eigenvalue of the Hamiltonian 
\begin{align}
H^{\rm SC}_0 \phi(\vec{r})\!=\!\varepsilon \phi(\vec{r}) \mbox{ where } H^{\rm SC}_0\!=\!\frac{\hat{p}_{z}^2}{2m^*}\!+\!\frac{\hat{p}_{r}^2}{2m^*}\!+\!\frac{\hat{p}_{\varphi}^2}{2m^*}\!-\!\mu^{(s)}\nonumber
\end{align}
The functions $\bar{G}$ and $\bar{F}$ are determined by the following equations:
\begin{align}
\bar{G}=\frac{1}{2\tau \bar{m}_J}\sum_{|m_J| < \bar{m}_J} \frac{i\omega_n+\bar{G}+H m_J}{\sqrt{(\Delta+\bar{F})^2-(i\omega_n+\bar{G}+H m_J)^2}}\\
\bar{F}=\frac{1}{2\tau \bar{m}_J}\sum_{|m_J| < \bar{m}_J} \frac{\Delta+\bar{F}}{\sqrt{(\Delta+\bar{F})^2-(i\omega_n+\bar{G}+H m_J)^2}}
\end{align}
with $\tau$ being the elastic scattering time and $\bar{m}_J \sim p_F R_3$ being the angular momentum cutoff. In the limit $H \rightarrow 0$, the leading order corrections to the above equations appear in quadratic order since linear terms vanish due the averaging over $m_J$. Indeed, one can show that the self-consistent solution for $\tau \rightarrow 0$ is given by
\begin{align}
\bar{G}&=\frac{i}{2\tau}\sin z \\
\bar{F}&=\frac{i}{2\tau}\cos z \\
\frac{\omega_n}{\Delta}& =\tan z-\kappa  \sin z
\end{align}
where $\kappa=3H^2 \tau \langle m_J^2 \rangle/\Delta$ is the characteristic scale for the magnetic field effects in the problem. Here $ \langle m_J^2 \rangle=1/ \bar{m}_J \sum_{|m_J| < \bar{m}_J} m_J^2 \sim (p_F R_3)^2$. Thus, corrections to the pairing gap are governed by the small parameter $\kappa \ll 1$. In terms of the flux quantum, this condition reads $\Phi/\Phi_0 \ll R_3/\xi_d$. Note that disorder suppresses orbital effects of the magnetic field and leads to a weaker dependence of the pairing gap on magnetic field (i.e., quadratic vs linear). In other words, the disordered superconductor can sustain much higher magnetic fields compared to the clean one, see Fig.\ref{fig:realistic_superconducting_shell}. Finally, the analysis above can be extended to $n\neq 0$. After some manipulations, one finds that \cite{Liu2001,Koshnick2007} 
\begin{align}
\frac{\Delta(\Phi)-\Delta_0}{\Delta_0} \sim \frac{\xi_d^2}{R_3^2}\left(n-\frac{\Phi}{\Phi_0}\right)^2
\end{align}
This estimate is consistent with the numerical calculations, see Fig. \ref{fig:realistic_superconducting_shell}.

\section{Derivation of the effective Hamiltonian.}
\label{app:effective}

In the previous section we derived the Green's function for the disordered superconducting ring. One can now use these results to study the proximity effect of the SC ring on the semiconducting core. We will focus here on the case when the SC shell is thin $d \sim l_e$ such that $\frac{\xi_d}{R_3} \ll 1$. In this case, one can neglect magnetic field dependence of the self-energy for the entire lobe. (Alternatively, when $\xi_d \sim R_3$ one can neglect magnetic field effect when $n-\frac{\Phi}{\Phi_0} \ll 1$). Thus, one can use zero field  Green's functions for the disordered superconductor to investigate the proximity effect which are obtained by substituting $\omega_n \rightarrow \tilde{\omega}_n = \omega_n \eta(\omega_n)$ and $\Delta_0 \rightarrow \tilde{\Delta}_0=\Delta_0 \eta(\omega_n)$ with $\eta(\omega_n)=1+1/{2\tau \sqrt{\omega_n^2+\Delta_0^2}}$ in the clean Green's functions. 

One can now integrate out the superconducting degrees of freedom and calculate the effective self-energy due to the tunneling between semiconductor and superconductor. Using the gauge convention when $\Delta_0$ is real, tunneling Hamiltonian between SM and SC is given by~\cite{Stanescu2011}
\begin{align}
H_t=\int dr dr' T(r,r')e^{i n\varphi/2} \Psi^\dag(r)\Psi(r')+H.c.
\end{align}
where $r$ and $r'$ refer to the SM and SC domains, respectively. $T(r,r')$ is the tunneling matrix element between the two subsystems, and $\Psi$ and $\Psi^\dag$ are the fermion annihilation and creation operators in the corresponding subsystem. One can calculate the SC self-energy due to tunneling to find
\begin{align}
\Sigma^{(\rm SC)}(r,\omega_n)=\Gamma(r)\frac{i \omega_n \tau_0 -\Delta_0 \left[\cos(n \varphi)
                     \tau_x\!+\!\sin(n \varphi)\tau_y\right]}{\sqrt{\omega_n^2+\Delta_0^2}}
                     \label{eq:selfenergy}
\end{align}
where $\Gamma(r)$ is a quickly decaying function away from $r=R_2$ describing tunneling between the two subsystems. Note that the SC self-energy in this approximation is the same as for a clean superconductor because the ratio of $\tilde{\omega}_n/\tilde{\Delta}_0$ is independent of $\tau$.

The Green's function for the semiconductor can be written as
\begin{align}\label{eq:Greens}
G^{-1}(\omega_n)=-i\omega_n-H_{\rm SM}-\Sigma^{(\rm SC)}(r,\omega_n) 
\end{align}
In order to calculate quasiparticle energy spectrum one has to find the poles of above Green's function. 

In the hollow cylinder limit, $\Gamma(r=R_2)$ is a constant and one can find low energy spectrum analytically. Indeed, after expanding Eq.~\eqref{eq:Greens} in small $\omega_n$, the quasiparticle poles are determined by the spectrum of the following effective Hamiltonian:  
\begin{align}
H_{\rm eff}=\frac{H_{\rm SM}}{1+\Gamma/\Delta_0}  &- \frac{\Gamma}{1+\Gamma/\Delta_0} \left[\cos(n \varphi)
                     \tau_x\!+\!\sin(n \varphi)\tau_y\right]=0
\end{align}
By comparison with Eq.~(2) of the main text, one can establish the correspondence between the renormalized and bare parameters of the semiconductor and proximity-induced gap $\Delta=\Delta_0 \Gamma /(\Delta_0+\Gamma)$.  

\section{Effect of higher $m_J$ states on the gap}
\label{app:highermJ}

As demonstrated in the main text, states with larger $m_J \neq 0$ have the potential to close the gap and thus limit the extent of the topological phase. Here we provide analytical estimates within the hollow cylinder model for the regions in parameter space that become gapless due to higher $m_J$ states. We start with the BdG Hamiltonian~(5) of the main text assuming $n=1$,
\begin{align}\label{eq:H_mj}
\tilde{H}_{m_J}&=
\left[\frac{p_{z}^2}{2m^*}-\mu_{m_J}\right]\!\tau_z+V_{Z}\sigma_z\\
&+A_{m_J}\! + C_{m_J}\sigma_z\tau_z+\alpha_2 p_z \sigma_y \tau_z+\Delta\tau_x,
\end{align}
with
\begin{eqnarray}
\mu_{m_J}&=&\mu-\frac{1}{8m^*R_2^2}\left(4m_J^2+1+\phi^2\right)-\frac{\alpha_1}{2R_2}\,,\\
    V_{Z}&=&\phi\,\left(\frac{1}{4m^*R_2^2}+\frac{\alpha_1}{2R_2}\right)\,,\\
    A_{m_J}&=&-\frac{\phi m_J}{2m^* R_2^2}\,,\\
    C_{m_J}&=&-m_J\left(\frac{1}{2m^*R_2^2}+\frac{\alpha_1}{R_2}\right)\,,
    \end{eqnarray}
with $\phi=1-\Phi(R_2)/\Phi_0$. With respect to the main text, we also introduced anisotropic spin-orbit coupling with $\alpha_1$ and $\alpha_2$ representing the strength of coupling to the transversal and longitudinal ($z$) momentum direction. In the main text, we used isotropic spin-orbit $\alpha_1=\alpha_2=\alpha$ but it is convenient for the discussion below to distinguish the two contributions.

\begin{figure}[t!]
\begin{center}
\includegraphics[width=\columnwidth]{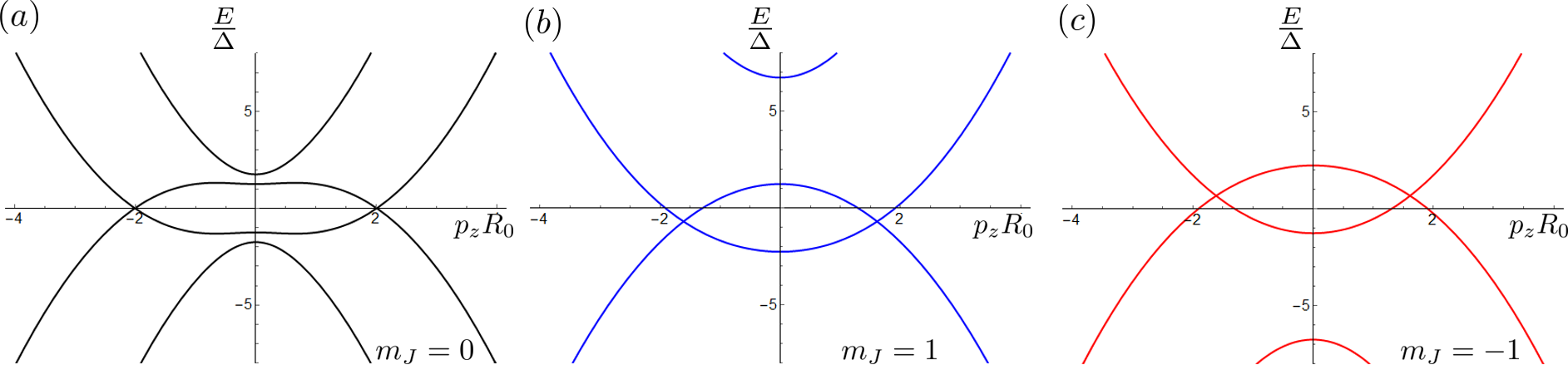}
\caption{Energy spectrum for the lowest $m_J$ sectors for $n=1$. Here parameters used are the same as in Fig.~2 of the main text, except $\Delta=0$. For finite $\Delta$ the intersections of particle and hole bands become avoided crossings. Note that for $m_J\neq 0$ these avoided crossings happen at finite energy, which leads to condition \eqref{cond_2a}.}
\label{fig:mJspectrum}
\end{center}
\end{figure}

Example energy spectra for the lowest $m_J$ sectors are shown in Fig.~\ref{fig:mJspectrum}. Particle-hole symmetry relates $m_J$ and $-m_J$ sectors. Therefore, $m_J=0$ sector is special in this sense. Note that $\alpha_2$ is crucial to estimate the topological gap in the $m_J=0$ sector, i.e., a topological gap requires $\alpha_2\neq 0$ and $\Delta\neq 0$. The conditions for a finite gap in $m_J\neq 0$ sectors are more stringent. First of all, the pairing term hybridizes states within each $m_J$ sector. Thus, the system is {\it gapless} if there is an odd number of particle and hole bands at the Fermi level, which leads to the  condition 
\begin{align}
(A_{mj}+V_Z\sigma_z)^2 & > \Delta^2+ (C_{mj}\sigma_z-\mu_{mj})^2\, , \label{cond_1}
\end{align}
which follows from the gap closing at $p_z=0$. The gapless region in the upper right corner of Figs.~2A and 2B of the main text is due $m_J=\pm 2$ states fulfilling condition \eqref{cond_1}.

If condition~\eqref{cond_1} is not satisfied and the number of bands at the Fermi levels is even, the system can be gapped -- see, for instance, panels (b) and (c) in Fig.~\ref{fig:mJspectrum}. This happens, for example, if the effective chemical potential for a given subband $\mu_{mj}-C_{mj}\sigma_z<0$ in which case the subband is empty and gapped. However, if the subband is filled, i.e. if $\mu_{mj}-C_{mj}\sigma_z>0$, one has to investigate the closing of the gap at finite momenta. In this case, the system is gapless when $\Delta$ is smaller than a certain critical value required to hybridize particle and hole bands with mismatched Fermi momenta, see Fig.~\ref{fig:mJspectrum}(b) and (c). In the limit $\alpha_2\rightarrow 0$, the condition for a {\it vanishing} gap reads
\begin{align}
|A_{mj}+V_Z\sigma_z| &> \Delta \label{cond_2a}\\
\mu_{mj}-C_{mj}\sigma_z &>0. 
\label{cond_2b}
\end{align}
One may notice that the term $A_{mj}+V_Z\sigma_z$ acts as a Pauli limiting field for a given subband and leads to pair-breaking effects. 
     
We can understand the generally finite extent of the gapped regions in the $\alpha$-$\mu$ plane by considering conditions \eqref{cond_2a} and \eqref{cond_2b}. Condition \eqref{cond_2a} is either met for sufficiently large $m_J$ or sufficiently large $\alpha_1$ (when $m_J$ is kept constant). At the same time, large $m_J$ states generally violate condition \eqref{cond_2b} since the bottom of the band is shifted up $\propto m_J^2$ which needs to be compensated by sufficiently large $\mu$. We therefore expect to find gapless states for large $\mu$ (which enable large $m_J$) or very large $\alpha$ which fulfill condition $\eqref{cond_2a}$ while still being compatible with condition $\eqref{cond_2b}$.

\section{Details about the numerical simulations of clean systems}
\label{app:clean_numerics_details}

From the numerical perspective, the solution of Eq.\eqref{eq:Greens} for the poles is not optimal given that one has to solve the non-linear equation for $\omega$. Therefore, we employ an alternative approach in which we couple SM to an artificial clean superconductor. We use the parameters for the superconductor and tunneling Hamiltonian such that in the end we reproduce Eq.~\eqref{eq:Greens} after integrating out the SC degrees of freedom.    
 
To obtain the correct self-energy Eq.~\eqref{eq:selfenergy}, the thickness of the simulated clean superconductor needs to be made significantly larger than the coherence length. This is achieved by using $R_3=3\,\mu$m for the simulations shown in Fig.~3 and 4 of the main text.
All parameters are chosen independent of
$r$ except of
\begin{equation}
  \begin{aligned}
    \Delta(r) &=\begin{cases}
      0 & r < R_2 \\
      \Delta & r \ge R_2\\
    \end{cases},\\
    \alpha &=\begin{cases}
      \alpha & r < R_2 \\
      0 & r \ge R_2\\
    \end{cases},\\
    A_\varphi &=\begin{cases}
      \Phi(R_2) r/(2 \pi R_2^2) & r < R_2 \\
      \Phi(R_2)/(2 \pi r) & r \ge R_2\\
    \end{cases}.
  \end{aligned}
\end{equation}
Here, $\Phi(R_2)$ corresponds to the flux penetrating the semiconducting core. In accordance to the arguments above, we simulate the superconductor without magnetic field. We solve Eq.~4 in the main text with the finite difference method, using a discretization length of 5\,nm, as detailed in Ref.~\cite{Winkler17}.

\section{Full cylinder semiconductor model in the small radius limit.}
\label{app:full_cylinder_analytics}

In this section we consider the full cylinder limit discussed in the main text ($R_1\rightarrow 0$ in Fig.~1). Using an effective model we demonstrate analytically the topological phase exists when exactly one superconducting flux quantum penetrates the core. The results of this section are complimentary to the numerical calculations of the main text. The effective Hamiltonian for the model is given by
\begin{widetext}
\begin{align}\label{eq:HBdGr11}
  \tilde{H}_{\rm BdG}&=\left(\frac{p_{z}^2}{2m^*}-\frac{1}{2m^* r} \frac{\partial }{\partial r} r \frac{\partial }{\partial r}-\mu\right)\tau_z +\frac{1}{2m^*r^2}\left(m_J-\frac{1}{2}\sigma_z-\frac{1}{2} n \tau_z + \frac{b}{2}\frac{r^2}{R_2^2}\tau_z\right)^2 \tau_z \nonumber \\ &- \frac{\alpha}{r} \sigma_z \tau_z \left(m_J-\frac{1}{2}\sigma_z-\frac{1}{2} n \tau_z + \frac{b}{2}\frac{r^2}{R_2^2}\tau_z\right)+\alpha p_z \sigma_y \tau_z
                       +\Delta(r)\tau_x.
\end{align}
and, unlike in the hollow cylinder limit, one has to solve the radial part of Eq. \eqref{eq:HBdGr11}.
We introduced the dimensionless variable $b=eB R_2^2=\pi B R_2^2/\Phi_0$.
The proximity-induced gap $\Delta(r)$ must vanish in the middle of the core, $\lim_{r\to 0}\Delta(r)= 0$.
We consider below the case when $\Delta(r)=\Delta r/R_2$, although the particular choice for the radial dependence of $\Delta(r)$ is not important for the demonstration of the existence of the topological phase.

We restrict our analysis to the $m_J=0$ sector for $n=1$ in the limit $1/m^* R_2^2 \gg \alpha/R_2$.
In this case, the problem at hand can be simplified since the Hamiltonian becomes separable at $\alpha\rightarrow 0$ and effect of spin-orbit can be included perturbatively.
In the limit $\alpha \rightarrow 0$, the electron spin is conserved and the Bogoliubov transformation diagonalizing Hamiltonian~\eqref{eq:HBdGr11} can written as
\begin{align}
\!\gamma_{\lambda, p_z, \sigma}\!=\!\!\int_0^{R_2}\!\!\! rdr \!\left[U_{\lambda, p_z, \sigma}(r)\Psi_{p_z, \sigma}(r) \!+\!V_{\lambda, p_z, -\sigma}(r)\Psi^\dag_{p_z, -\sigma}(r) \right]\,,
\end{align}
where the transformation coefficients $U_{\lambda, p_z, \sigma}(r)$ and $V_{\lambda, p_z, \sigma}(r)$ are given by the solution of Eq.~\eqref{eq:HBdGr11}.
Neglecting the spatial dependence of $\Delta(r)$, the functions $U_{\lambda, p_z, \sigma}(r)$ and $V_{\lambda, p_z, \sigma}(r)$ can be approximately written as
\begin{align}
U_{\lambda, p_z, \sigma}(r)=u_{\lambda, \sigma}(p_z) f_{\lambda, \sigma}(r)\\
V_{\lambda, p_z, \sigma}(r)=v_{\lambda, \sigma}(p_z) f_{\lambda, \sigma}(r)\
\end{align}
where the single-particle wave functions $f_{\lambda, \sigma}(r)$ are defined by the following radial Schr\"{o}dinger equation:
\begin{align}\label{eq:radial}
-\frac{1}{2m^*}\left(\frac{1}{r} \frac{\partial }{\partial r} r \frac{\partial }{\partial r}\!-\!\frac{1 + \sigma_z}{2r^2}-\frac{b^2}{4R_2^2}\frac{r^2}{R_2^2}+\frac{b}{R_2^2}\frac{1+\sigma_z}{2}\right)f_{\lambda, \sigma}(r)=\varepsilon_{\lambda, \sigma}\,f_{\lambda, \sigma}(r)\,.
\end{align}
\end{widetext}
The linear term in $b$ represents a constant energy shift,
\begin{equation}
\delta_{\sigma} =
\begin{cases}
\dfrac{b}{2m^*R_2^2} & \sigma=\uparrow\,, \\ 0 & \sigma=\downarrow\,.
\end{cases}
\end{equation}
After introducing the dimensionless coordinate $x=r/R_2$ and the dimensionless energies $\kappa_{\lambda,\sigma}=2m^*R_2^2 (\varepsilon_{\lambda,\sigma}+\delta_{\sigma})$, the above equation becomes 
\begin{align}
\left(-\frac{1}{x} \frac{\partial }{\partial x} x \frac{\partial }{\partial x}\!+\!\frac{1+\sigma_z}{2x^2}+\frac{b^2}{4}x^2\right)f_{\lambda, \sigma}(x)=\kappa_{\lambda, \sigma}f_{\lambda, \sigma}(x)\,.
\end{align}
The normalized eigenstates of this equation, satisfying the boundary condition $f_{\lambda, \sigma}(x=1)=0$, are
\begin{align}
f_{\lambda, \uparrow}(r)&=C_{\lambda \uparrow}\,R_2^{-1}\, x e^{-x^2/4}  \, _{1}F_{1}\left(1-\frac{\kappa_{\lambda \uparrow}}{2b},2,\frac{x^2}{2}\right)\,,\\
f_{\lambda, \downarrow}(r)&=C_{\lambda \downarrow}\,R_2^{-1}\,e^{-x^2/4}  \, _{1}F_{1}\left(\frac{1}{2}-\frac{\kappa_{\lambda \downarrow}}{2b},1,\frac{x^2}{2}\right)\,.
\end{align}
Here, $_{1}F_{1}$ is the is the Kummer confluent hypergeometric function and the coefficients $C_{\lambda \sigma}$ are determined by the normalization condition
\begin{align}
\int_0^{R_2} |f_{\lambda, \sigma}(r)|^2\,rdr=1\,.
\end{align}
The corresponding eigenvalues are
\begin{align}
\varepsilon_{\lambda, \sigma}=\frac{\kappa^2_{\lambda \sigma}}{2 m^* R_2^2}-\delta_\sigma
\end{align}
where $\kappa_{\lambda,\sigma}$ are zeros of the appropriate Kummer confluent hypergeometric function for the two spins.
Taking all into account, for $b=1$ and $n=1$ the lowest eigenvalues of Eq.~\eqref{eq:radial} are
\begin{align}
\varepsilon_{1, \uparrow}\approx \frac{13.77}{2 m^* R_2^2}\,,\;\varepsilon_{1, \downarrow}\approx \frac{5.84}{2 m^* R_2^2}\,, \\
\varepsilon_{2, \uparrow}\approx \frac{48.30}{2 m^* R_2^2} \,,\; \varepsilon_{2, \downarrow}\approx \frac{30.55}{2 m^* R_2^2}\,.
\end{align}
Note that different values of $b$ will affect the numerical coefficients reported above.

In the limit $1/2m^* R_2^2 \gg \alpha/R_2, \Delta$, one can project the system to the lowest energy manifold (i.e. $\lambda=1$) and integrate over radial coordinate. After some algebra, the effective Hamiltonian takes the simple form (up to a constant):
\begin{align}\label{eq:HBdGeff1}\nonumber
\tilde{H}_{\rm BdG}&=\left(\frac{p_{z}^2}{2m^*}-\tilde{\mu}\right)\,\tau_z +\tilde{V}_Z \sigma_z + \tilde{\alpha}\,p_z \sigma_y \tau_z +\tilde{\Delta}\tau_x
\end{align}
where the effective parameters are given by:
\begin{align}
\tilde{\mu}&=\mu-\frac{\varepsilon_{1, \uparrow}\!+\!\varepsilon_{1, \downarrow}}{2}-\frac{\alpha}{2R_2}\,\left(A_\uparrow - \,\frac{B_\uparrow-B_\downarrow}{2}\right)\\
\tilde{V}_Z&=\frac{\varepsilon_{1, \uparrow}-\varepsilon_{1, \downarrow}}{2}+\frac{\alpha}{2R_2}\,\left(A_\uparrow-\,\frac{B_\uparrow+B_\downarrow}{2}\right) \\
\tilde{\alpha}&=\alpha\,C\,,\\
\tilde{\Delta}&=\Delta\,D\,.\label{eff_def}
\end{align}
with numerical constants $A_\sigma$, $B_\sigma$, $C$, $D$ given in terms of the overlap integrals:  
\begin{align}
A_\sigma&=\int_0^1 \left|f_{1,\sigma}(x)\right|^2 dx= \begin{cases} 2.056\dots & \sigma=\uparrow \\ 3.521\dots & \sigma=\downarrow \end{cases}\,,\\
B_\sigma&=\int_0^1 x^2 \left|f_{1,\sigma}(x)\right|^2 dx= \begin{cases} 0.552 \dots & \sigma=\uparrow \\ 0.423\dots & \sigma=\downarrow \end{cases}\,,\\
C&=\int_0^1 f_{1,\uparrow}(x)f_{1,\downarrow}(x) dx = 0.93\dots\,, \\
D&=\int_0^1 x^2 f_{1,\uparrow}(x)f_{1,\downarrow}(x) dx = 0.465\dots\,.
\end{align}

One can notice that the Zeeman term remains finite at $b=1$ (i.e. one flux quantum) in contrast to the hollow cylinder model.
As mentioned in the main text, this is because the semiconducting states are distributed through the semiconducting core rather than localized at $r=R_2$, so that the flux cannot perfectly cancel the effect of the winding number.

In summary, we have shown that full cylinder model also maps onto Majorana nanowire model of
Refs.~\cite{Lutchyn10, Oreg10} and supports topological superconducting phase. The topological quantum phase transition from the topologically trivial (i.e. s-wave) to non-trivial (i.e. p-wave) phases occurs at 
\begin{align}
|\tilde{V}_Z|=\sqrt{\tilde{\mu}^2+\tilde{\Delta}^2}.
\end{align}
Note that so far we have considered $m_J=0$ sector. One needs to investigate other $m_J$ sectors and make sure that quasiparticle gap does not close in the topological phase. This can be done numerically, see main text.

\begin{figure}[h]
\centering
\includegraphics[width=3in]{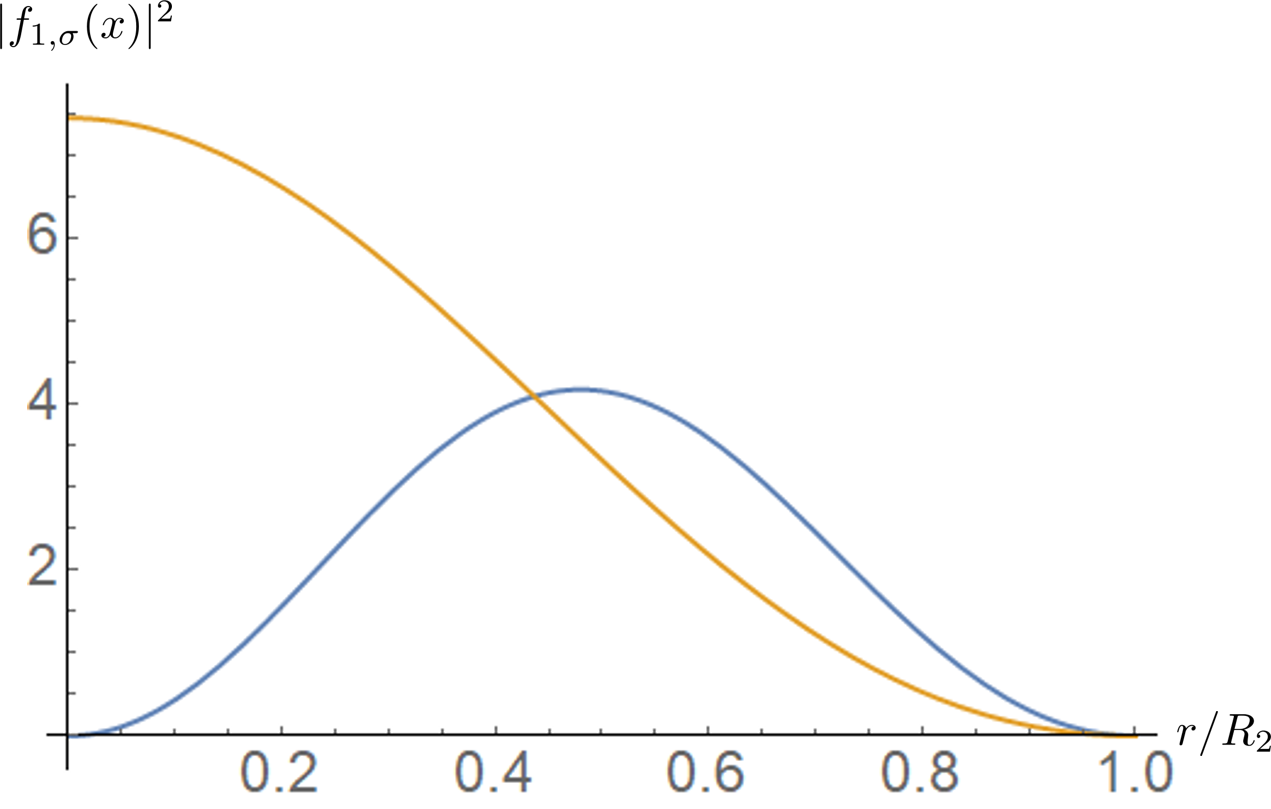}
\caption{(Color online). Probability density for the lowest-energy spin-up (blue) and spin-down (yellow) modes.}
\label{fig:wavefunctions}
\end{figure}

\section{Additional DOS plots for the full cylinder model}
\label{app:additional_dos}

\begin{figure}
  \begin{center}
    \includegraphics[width=\columnwidth]{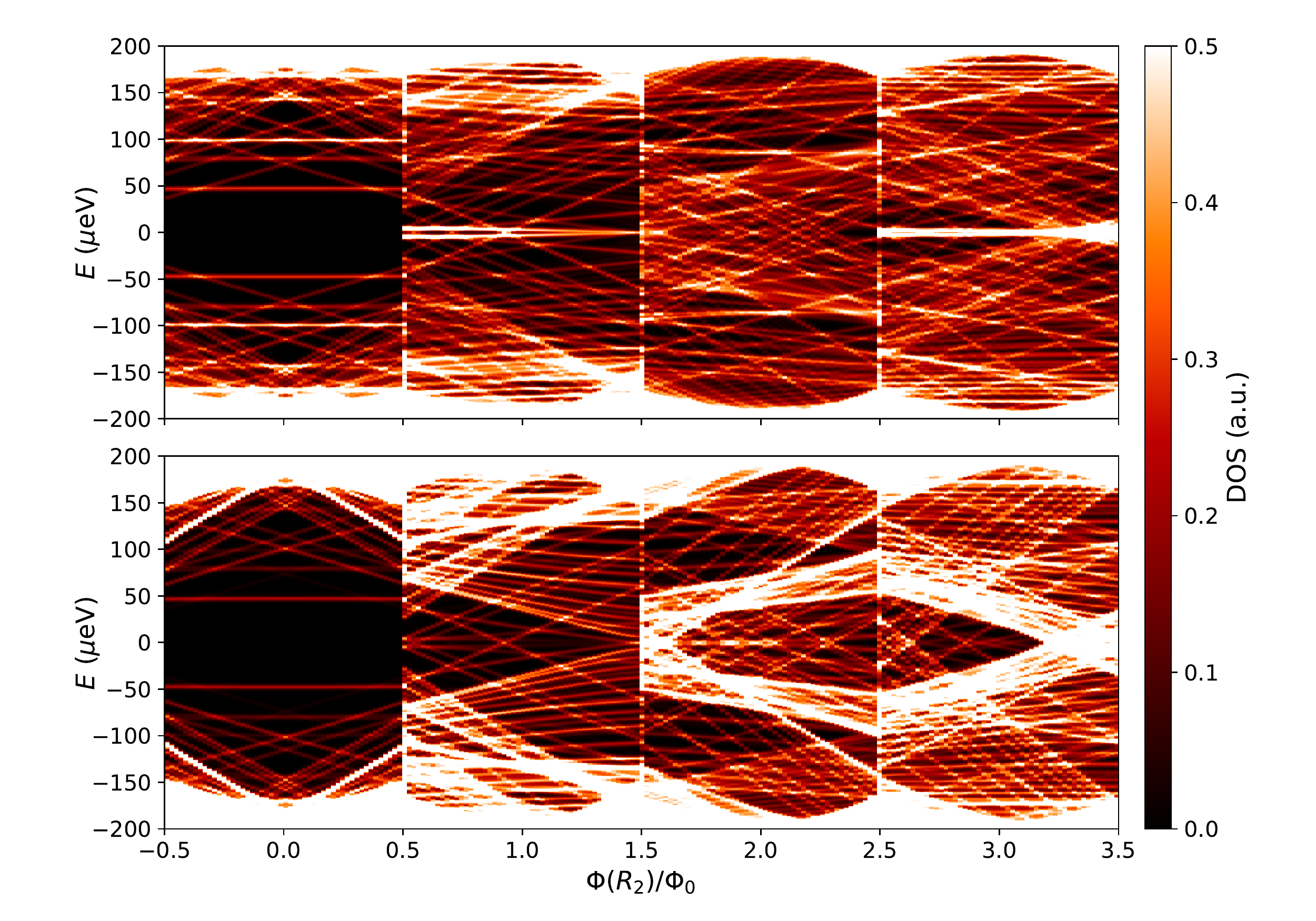}
\caption{DOS at the end (top panel) and in the middle (bottom panel) of a finite wire of 3\,$\mu$m length as a function of flux for $\mu=1.5$\,meV, 
  $\alpha=40$\,meV\,nm and the same parameters as in Fig.~2~{\bf A}.
  \label{fig:little_parks_addition}}
\end{center}
\end{figure}

In order to understand the topological phase transition within as a function of magnetic flux within the same lobe, it's useful to study the bulk DOS calculated, for example, in the middle of the wire. We focus on $n=1$ lobe in Fig.~\ref{fig:little_parks_addition}~(a) where the topological phase transition manifests itself by closing of the bulk. It's also enlightening to compare the bulk DOS and local DOS at the ends of the wire shown in Fig.~3~\textbf{B} of the main text. One may notice the asymmetry with respect to the center of the lobes which follows from the different dependence of the semiconducting and superconducting states on magnetic field.      

This asymmetry depends on parameters and in Fig.~\ref{fig:little_parks_addition}~(b) we show the boundary DOS for a different set of parameters, in which the zero bias peaks extend throughout the entire $n=1$ and $n=3$ lobes. Note, however, that according to Fig.~3~\textbf{A} the system is gapless for this parameters at $\Phi(R_2)=\Phi_0$. However, as discussed in the main text, the rotational-symmetry-breaking perturbations (e.g. disorder) may lead to gap opening for $m_J\neq 0$ states and therefore stabilize the topological phase.

\section{Details about the numerical simulations of disordered systems}
\label{app:disorder_numerics_details}

\begin{figure}
  \begin{center}
    \includegraphics[width=\columnwidth]{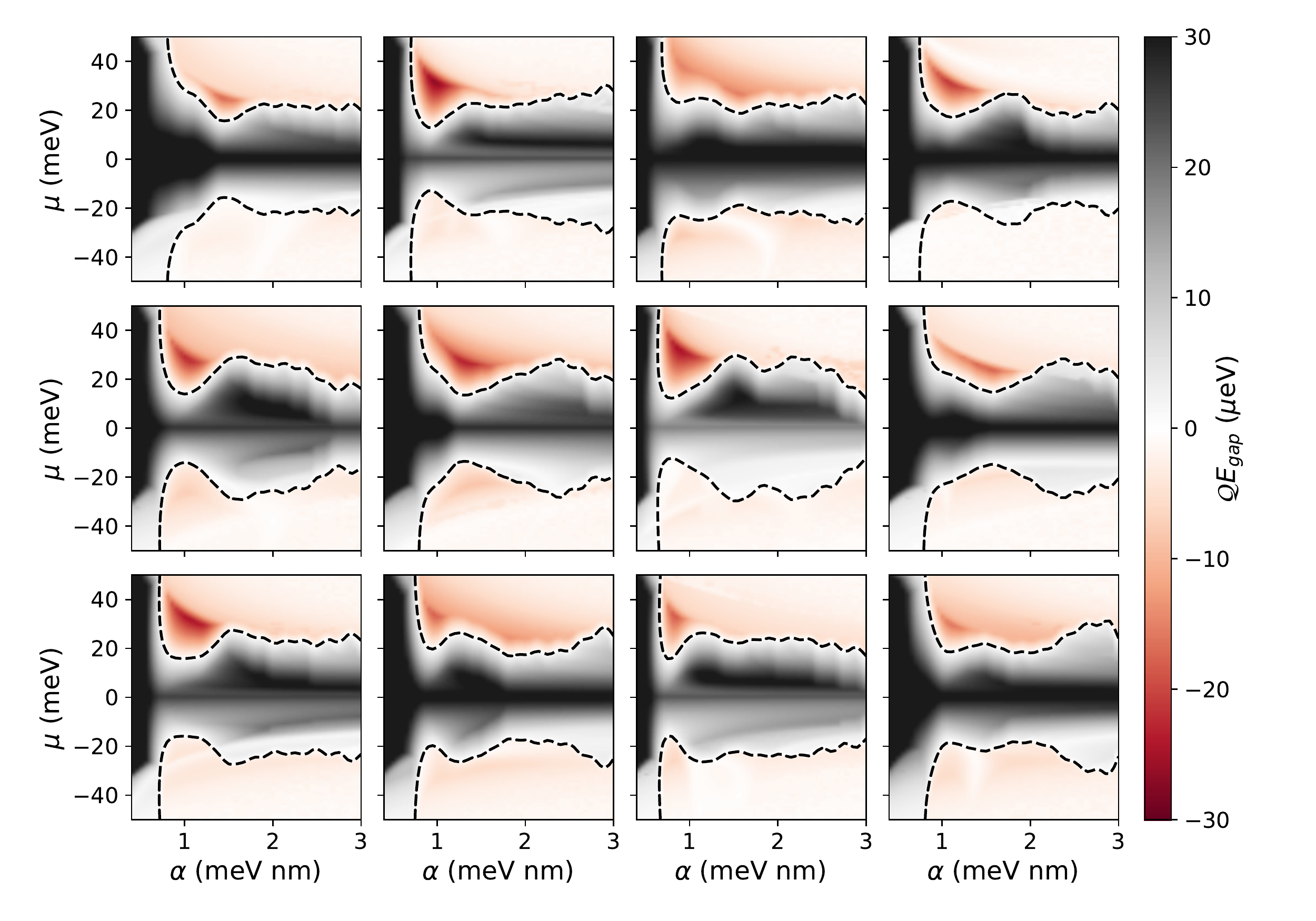}
\caption{Phase diagrams for different disorder realizations used to
  obtain the average in
  Fig.~3. 
  \label{fig:disordered_realizations}}
\end{center}
\end{figure}

For the simulations with disorder in Fig.~4 of the main text, we use the Kwant package~\cite{kwant} for discretizing the Hamiltonian on a 2D square lattice, using a lattice spacing of 10\,nm. The system is assumed to be translation-invariant along the $z$ direction, with the disorder configuration repeating along the $z$-axis. This trick is required since a full 3D simulation would be computationally too demanding. To accommodate to the higher computational cost we use a smaller $R_3$ of $1.5\,\mu$m in these simulations. In Fig.~\ref{fig:disordered_realizations} we show phase diagrams for different disorder realizations.

\bibliography{synopsis}
\end{document}